\documentclass[prd,twocolumn,nofootinbib,showpacs,eqsecnum,floatfix,superscriptaddress]{revtex4}
\usepackage{latexsym,amsbsy}
  \usepackage[dvips]{graphicx}
 \DeclareGraphicsExtensions{.eps, .jpg}

\begin{document}
\title{Multidimensional gravitational model with anisotropic pressure
 }
\author{O. A. Grigorieva, G. S. Sharov}
\email{german.sharov@mail.ru} \affiliation{Tver state university,
170002, Sadovyj per. 35, Tver, Russia}
\date{\today}

\begin{abstract}
We consider the gravitational model with additional spatial
dimensions and anisotropic pressure which is nonzero only in these
dimensions. Cosmolo\-gical solutions 
in this model include accelerated expansion of the Universe at
late age of its evolution and dynamical compactification of extra
dimensions. This model describes observational data for Type Ia
supernovae on the level or better than the $\Lambda$CDM model. We
analyze two equations of state resulting in different predictions
for further evolution, but in both variants the acceleration epoch
is finite.

\end{abstract}

\maketitle

\section{Introduction}\label{Intr}

The most important event of last 15 years in astrophysics is
conclusion about accelerated expansion of our universe at late
stage of its evolution. This conclusion was based on observations
of luminosity distances and redshifts for the Type Ia supernovae
\cite{accPerl,Riess}, cosmic microwave background \cite{WMAP},
large-scale galaxy clustering \cite{SDSS}, and other evidence
\cite{CopelandST06,Clifton}.

To explain accelerated evolution of the universe various
mechanisms have been suggested, including the most popular
cosmological model $\Lambda$CDM with a $\Lambda$ term (dark
energy) and cold dark matter (see reviews
\cite{CopelandST06,Clifton,BambaCapNO12,Kunz12}). The $\Lambda$CDM
model with 4\% fraction of visible (baryonic) matter nowadays,
23\% fraction of dark matter and 73\% fraction of dark energy
\cite{WMAP} describes Type Ia supernovae, data rather well and
satisfies observational evidence, connected with rotational curves
of galaxies, galaxy clusters and anisotropies of cosmic microwave
background. However, the $\Lambda$CDM model (along with vague
nature of dark matter and energy) has some problems with fine
tuning of the observed value of $\Lambda$, which is many orders of
magnitude smaller than expected vacuum energy density, and with
different time dependence of dark energy $\Omega_\Lambda$ and
material $\Omega_m$ fractions (we have
$\Omega_\Lambda\simeq\Omega_m$ nowadays).

Therefore a large number of alternative cosmological models have
been proposed. They include theories with extra dimensions
\cite{Sahdev,PahwaChS,Mohammedi02,Darabi03,Darabi09,BringmannEG03,PanigrahiZhCh06,MiddleSt11,FarajollahiA10,Neupane,Qiang05,Leon10,MaartensK};
matter with nontrivial equations of state, for example, Chaplygin
gas \cite{KamenMP01,GKamenMP03}; scalar fields with a potential
\cite{CaldwellDS98,ArmenMukhS00,Chameleon03}; modified gravity
with $f(R)$ Lagrangian \cite{SotiriouF,NojOdinFR} and many others
\cite{Clifton,CopelandST06,BambaCapNO12,Kunz12}.

 In this paper we explore the
 cosmological model with anisotropic pressure and
nontrivial equation of state in $1+3+d$ dimensions, suggested by
Pahwa, Choudhury and Seshadri in Ref.~\cite{PahwaChS}. The authors
omitted the important case $d=1$, we include it into
consideration. We also analyze how to modify the equation of state
and to avoid ``the end of the world'' (the finite-time future
singularity) which is inevitable in the model \cite{PahwaChS}.

In this model the $1+3+d$ dimensional spacetime 
is symmetric and isotropic in two subspaces: in 3 usual spatial
dimensions and in $d$ extra dimensions. It has the following
metric with two Robertson--Walker terms \cite{PahwaChS}:
 \begin{eqnarray}
 ds^2 &=& -dt^2+a^2(t)\left(\frac{dr^2}{1-k_1 r^2} + r^2
 d\Omega\right)\nonumber\\
 &+&b^2(t)\left(\frac{d R^2}{1-k_2 R^2}+R^2 d\Omega_{d-1}
\right).
 \label{metricInd}
 \end{eqnarray}
 Here the signature is $(-,+,\dots,+)$, the speed of
light $c=1$, $a(t)$ and $k_1$ is are the scale factor and
curvature sign in usual dimensions, $b(t)$ and $k_2$ are
corresponding values for extra dimensions. It is supposed in
Ref.~\cite{PahwaChS} that the scale factor $a(t)$ grows while
$b(t)$ diminishes, in other words, some form of dynamical
compactification
\cite{PahwaChS,Mohammedi02,Darabi03,Darabi09,BringmannEG03,PanigrahiZhCh06,MiddleSt11,FarajollahiA10,Neupane,Qiang05,Leon10}
takes place, a size of compactified $b$ is small enough to play no
essential role at the TeV scale.

The authors of Ref.~\cite{PahwaChS} develop the approach of
Ref.~\cite{Sahdev} and suppose that the spacetime
(\ref{metricInd}) is filled with a uniform density matter with
anisotropic pressure and the following energy-momentum tensor:
\begin{equation}
 T^{\mu}_{\nu} = \mbox{diag}\,(-\rho,P_a,P_a,P_a,P_b,\dots,P_b). \label{Tmn}
\end{equation}
 Here $\rho$ is the energy density  and $P_a$ ($P_b$) is
the pressure in normal (extra) dimensions.  So in normal
dimensions pressure is different from that in additional
dimensions, while being isotropic within each subspace.

In  Ref.~\cite{PahwaChS}  matter in the form of  a single fluid
 is supposed to behave like pressureless dust
($P_a=0$) in usual dimensions, while in extra dimensions it has
appreciable pressure $P_b$ depending on density $\rho$ by a power
law
\begin{equation}
P_a=0,\qquad P_b=W\rho^{1-\gamma}
 \label{Pbrho1}
\end{equation}
 with a negative constant $W$. The latter equation of state resembles
a generalized Chaplygin gas \cite{GKamenMP03}. In this model
matter (\ref{Tmn}) with anisotropic pressure plays a role of dark
energy and source of accelerated expansion. So the following
Einstein equation without usual $\Lambda$ term  is considered:
 \begin{equation}
 G^\mu_\nu=8\pi G T^\mu_\nu.
 \label{Eeq}\end{equation}

To describe the late time acceleration of the universe many
authors
\cite{Sahdev,PahwaChS,Mohammedi02,Darabi03,Darabi09,BringmannEG03,
PanigrahiZhCh06,MiddleSt11,FarajollahiA10} used the similar
approach, in particular, extra dimensions, a metric of the type
(\ref{metricInd}) and the energy-momentum tensor (\ref{Tmn}).
However, the cited authors used different equations of state. In
particular, in
Refs.~\cite{BringmannEG03,PanigrahiZhCh06,MiddleSt11} these
equations were linear
 \begin{equation}
P_a=w_a\rho,\qquad P_b=w_b\rho.
 \label{EoSl}
\end{equation}
 Under these conditions a set of cosmological solutions with power law dependence
of $a$, $b$, $\rho$ on $t$ was obtained in
Refs.~\cite{BringmannEG03,PanigrahiZhCh06}. But for these
solutions an acceleration for $a$ and a dynamical compactification
or stabilization for $b$ are not possible simultaneously. The
similar problem appears in Ref.~\cite{FarajollahiA10}, where the
authors use the sum of two perfect fluids with densities $\rho$
and $\bar\rho$ and the equations of state $P_a=w_a\rho$,
$P_b=w_b\bar\rho$. In this case for solutions with $a\sim
t^\alpha$ an acceleration ($\alpha>1$) suppresses any
compactification or diminishing for $b(t)$.

The problem of dynamical compactification for the extra dimensions
was solved in the paper  by Mohammedi \cite{Mohammedi02} under
assumptions (\ref{metricInd}) with $k_2=0$, (\ref{Tmn}) and the
following ansatz:
 \begin{equation}
b=\mbox{const}\cdot a^{-n}.
 \label{Mohans}
\end{equation}
 Mohammedi constructed solutions with accelerated expansion
 without a predetermined equation of state.  In
his approach evolution of values $\rho$, $P_a$, $P_b$ was
calculated from the right hand sides if Eqs.~(\ref{Eeq}) with  a
$\Lambda$ term. Relations between these values correspond to
equations of state, they appear at the last stage of this scheme.
Application of the Mohammedi's solutions \cite{Mohammedi02} to
describing the observational data will be discussed below.

Middleton and Stanley in Ref.~\cite{MiddleSt11} in the framework
of the linear equations of state (\ref{EoSl}) deduced the relation
 $$b= a^{-n}\Big(C_1+C_0\int a^{n-3}dt\Big),$$
generalizing Eq.~(\ref{Mohans}). Here  $n=(3w_a-2w_b-1)/(1-w_b)$.
They obtained  a set of cosmological solutions  including a
hypergeometric function of powers of $a$. However, for these
solutions an accelerated expansion of $a$ takes place only when
the EoS parameters $w_a$, $w_b$ in Eq.~(\ref{EoSl}) are both
negative, and also an accelerated expansion of $a$ in the late
universe is incompatible with dynamical compactification of $b$
\cite{MiddleSt11}. This conclusion corresponds to the findings in
Refs.~\cite{BringmannEG03,PanigrahiZhCh06}.

It is worth noting that the cosmological acceleration with the
dynamical compactification of extra dimensions may be achieved in
scalar-tensor theories, in particular, in 5-dimensional
Brans-Dicke models \cite{Qiang05,Leon10}. But these models along
with the extra metric component $g_{44}$ require the additional
degree of freedom in the form of the scalar Brans-Dicke field
$\phi$.

This paper is organized as follows. In Sec.~\ref{Cosm}  we show,
that the model \cite{PahwaChS} not only for $d\ge2$ but also in
the case $d=1$ can describe the current acceleration of the
universe with dynamical compactification of $b$. In Sec.~\ref{Obs}
we apply this model for all $d\ge1$ to describing observational
data for Type Ia supernovae and determine optimal model
parameters. In Sec.~\ref{Modif} we modify the model
\cite{PahwaChS} to solve the above mentioned problem of ``the end
of the world''.

\section{Cosmological solutions 
}\label{Cosm}

For the considered metric (\ref{metricInd}) in the case $k_2=0$
the Einstein tensor components $G^\mu_\nu$ ($\mu , \nu
=0,1,\dots,d+3$, $1\le i\le3<I$) are \cite{PahwaChS}:
\begin{eqnarray}
G^{0}_{0}&=&-3 d\frac{\dot{a}}{a}\frac{\dot{b}}{b}-
3\frac{\dot{a}^2}{a^2}-\frac{d(d-1)}{2} \frac{\dot{b}^2}{b^2}-3\frac{k_1}{a^2}, \nonumber \\
G^{i}_{i}&=&-2\frac{\ddot{a}}{a}-d\frac{\ddot{b}}{b}-2
d\frac{\dot{a}}{a}\frac{\dot{b}}{b}-\frac{\dot{a}^2}{a^2}-\frac{d(d-1)}{2}\frac{\dot{b}^2}{b^2}
-\frac{k_1}{a^2}, \nonumber \\
G^{I}_{I}&=&(1-d)\bigg[\frac{\ddot{b}}{b}+3\frac{\dot{a}}{a}\frac{\dot{b}}{b}+
\Big(\frac{d}{2}-1\Big)\frac{\dot{b}^2}{b^2}\bigg]-3\frac{\ddot{a}a+\dot{a}^2+k_1}
{a^2}. \nonumber
\end{eqnarray}

If we substitute these expressions into Eq.~(\ref{Eeq}) and add
the continuity condition $T^{\mu}_{\nu ; \mu}=0$ we obtain the
system of cosmological equations. This system has the form
\begin{eqnarray}
\frac{\dot{a}^2}{a^2}+\frac{\dot{a}}{a}\frac{\dot{b}}{b}+\frac{k_1}{a^2}
&=&\frac{8\pi G}3\rho,\label{e00}\\
2\frac{\ddot{a}}{a}+\frac{\ddot{b}}{b}+2\frac{\dot{a}}{a}\frac{\dot{b}}{b}+
\frac{\dot{a}^2}{a^2}+\frac{k_1}{a^2}&=&0,\label{e11}\\
-\frac{\ddot{a}a+\dot{a}^2+k_1}{a^2}&=&\frac{8\pi G
}3P_b,\label{e44}\\
 \frac{d}{dt}(\rho a^3 b)
+P_b a^3 \frac{d}{dt}b&=&0.  \label{contin}
\end{eqnarray}
in the case with $d=1$ extra spatial dimension, that did not
considered in Ref.~\cite{PahwaChS}. Here pressure $P_a$ in
``usual'' dimension equals zero, as mentioned above.
Eq.~(\ref{contin}) is the continuity condition for $d=1$ and
$P_a=0$.

Using the Hubble constant
 $H_0\simeq2.28\cdot10^{-18}$ c${}^{-1}$ \cite{WMAP} �
 and the critical density
  \begin{equation}
 \rho_c=\frac{3H_0^2}{8\pi G}
 \label{rocr}\end{equation}
  at the present time, we make the following substitutions
  \begin{equation}
\tau=H_0t,\!\quad\bar{\rho}=\frac{\rho}{\rho_c},\!\quad
\bar{p}_b=\frac{P_b}{\rho_c},\!\quad A=\log\frac a{a_0},\!\quad
B=\log \frac b{b_0}
 \label{tau} \end{equation}
  and introduce dimensionless time $\tau$, density
$\bar{\rho}$, pressure $\bar{p}_b$ and logarithms $A$, $B$ of the
scale factors (here $a_0$, $b_0$ are present time values of $a$
and $b$).

 We denote derivative with
respect to $\tau$ as primes and rewrite the system
(\ref{e00})~--~({\ref{contin}}) as follows:
 \begin{eqnarray}
{A'}^2+A'B'-\Omega_k e^{-2A}&=&\bar{\rho},\label{A1} \\
2A''+3{A'}^2+B''+{B'}^2+2A'B'&=&\Omega_k e^{-2A}, \label{A2}\\
A''+2{A'}^2-\Omega_k e^{-2A}&=&-\bar{p}_b, \label{A3}\\
\bar{\rho}'+3\bar{\rho}A'+(\bar{\rho}+\bar{p}_b)\,B'&=&0.
\label{rho}
 \end{eqnarray}
 Here
  \begin{equation}
 \Omega_k=-k_1(a_0H_0)^{-2}.
  \label{Omk} \end{equation}
If we express
 \begin{equation}
 B'=(\bar{\rho}+\Omega_ke^{-2A})/A'-A'
 \label{B} \end{equation}
 from Eq.~(\ref{A1}) and substitute it into three equations
(\ref{A2})~--~(\ref{rho}), one should note that Eq.~(\ref{A2}) may
be reduced to Eq.~(\ref{A3}). So in the planar case
$$k_1=k_2=0,\qquad\Omega_k=0$$
we have the system of two independent equations
 \begin{eqnarray}
A''&=&-2{A'}^2-\bar{p}_b,  \label{sysA} \\
\bar{\rho}'&=&-3\bar{\rho}A'+(\bar{\rho}+\bar{p}_b)
(A'-\bar{\rho}/A'). \label{sysrho}
 \end{eqnarray}

 If we fix an  equation of state for pressure $\bar{p}_b$, for example, the above mentioned power law
(\ref{Pbrho1})
 \begin{equation}
 \bar{p}_b=w{\bar{\rho}}^{\,1-\gamma},
 \label{pbro} \end{equation}
 we may consider the equations (\ref{sysA}), (\ref{sysrho}) as
a closed system of first order differential equations with respect
to 2 unknown functions $A'(\tau)$ and $\bar{\rho}(\tau)$. The
dependence (\ref{pbro}) is used in Ref.~\cite{PahwaChS}, where
parameters $w$ and $\gamma$ are chosen in accordance with
observations.

The Cauchy problem for the system (\ref{sysA}), (\ref{sysrho})
requires two initial conditions. We refer them to the present
epoch (here and below it corresponds to the value $\tau=1$) in the
following form:
 \begin{equation}
 A'\big|_{\tau=1}=1,\qquad \bar{\rho}\big|_{\tau=1}=\Omega_0.
 \label{initial} \end{equation}
 The first condition results from definition of the Hubble constant
  $$\;H_0=\frac{\dot{a}}a  
  \Big|_{t=t_0}=H_0 A'\big|_{\tau=1}.$$
 In the second condition (\ref{initial}) we suppose that the energy
density $\rho=\bar{\rho}\cdot\rho_c$ at the present time has the
fraction $\Omega_0$ in the critical density (\ref{rocr}). In
Ref.~\cite{PahwaChS} this fraction equals matter density fraction
in the $\Lambda$CDM model \cite{Clifton}:
 \begin{equation}
  \Omega_0=\Omega_m=0{.}27.
  \label{Omega0} \end{equation}
 Note that in Ref.~\cite{PahwaChS} the second condition (\ref{initial})
was used in the form  $\bar{\rho}\big|_{\tau=1}=1$, but the value
$\Omega_0$ (\ref{Omega0}) was taken as the factor in the r.h.s. of
Eq.~(\ref{Eeq}). From our point of view, that approach introduces
useless vagueness in physical sense of the value $\rho$. In our
approach  $\rho$ in conditions (\ref{initial}) is density of all
gravitating matter (visible and dark) with described above
anisotropic pressure.

Remind that we have no dark energy or $\Lambda$ term in
Eq.~(\ref{Eeq}) in the model \cite{PahwaChS}. Anisotropic pressure
in additional dimensions plays here the role of dark energy as a
source of acceleration. The contribution of this source is the
term $\Omega_B=-B'\big|_{\tau=1}$ in the equality
 \begin{equation}
  \Omega_m+\Omega_B+\Omega_k=1,
  \label{Omega3} \end{equation}
 that results from equation (\ref{A1}), if we fix it at the present
time $\tau=1$.

To obtain cosmological solutions for $d=1$, $k_1=0$ in this model
we are to solve numerically the Cauchy problem for the system
(\ref{sysA}), (\ref{sysrho}) with initial conditions
(\ref{initial}) moving into the past for $\tau<1$ and into the
future for $\tau>1$. Then we integrate functions $A'(\tau)$ and
$B'(\tau)$ (\ref{B}) keeping in mind Eqs.~(\ref{tau}) and
calculate dependence of the scale factors $a=a_0e^A$, $b=b_0e^B$
and density $\bar{\rho}$ on dimensionless time $\tau$.

\begin{figure}[th]
\includegraphics[scale=0.8,trim=2mm 1mm 2mm -1mm]{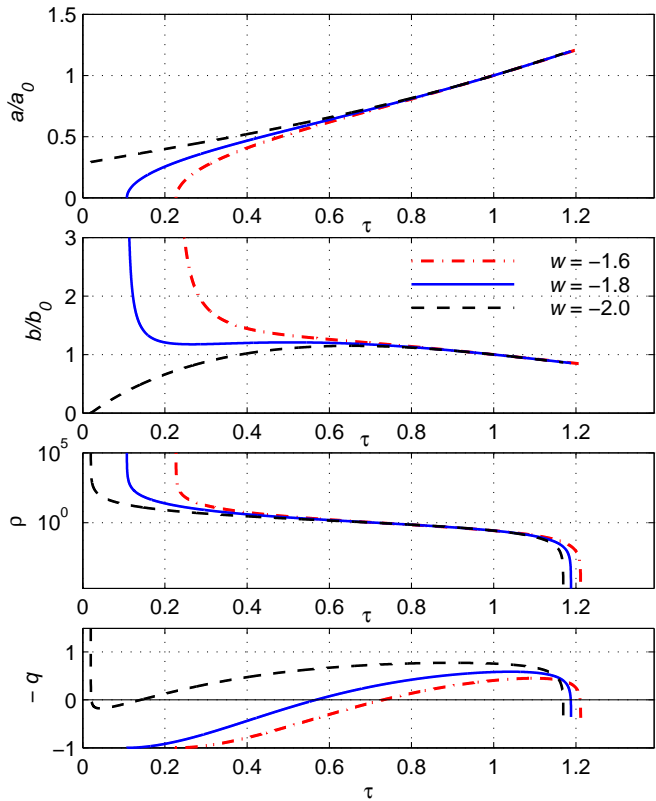}
\caption{Scale factors $a$,$ b$, density $\bar\rho$ and
acceleration parameter $-q$ depending on dimensionless time $\tau$
for $\Omega_0=0{.}27$, $\gamma=0{.}9$ and specified values of $w$}
 \label{If1}\end{figure}

The results of calculation for scale factors $a(\tau)$, $b(\tau)$,
density $\bar{\rho}(\tau)$ and the acceleration parameter
  \begin{equation}
 -q=\frac{\ddot a a}{{\dot a}^2}=\frac{A''+{A'}^2}{{A'}^2}
 \label{accel} \end{equation}
($q$ is the deceleration parameter) are presented in
Fig.~\ref{If1}. Here
 $k_1=0$, $\Omega_0=0{.}27$, $\gamma=0{.}9$ and
3 scenarios for $w=-1{.}6$ (dash-dotted line), $w=-1{.}8$ (solid
lines) and $w=-2$ (dashed lines) are shown.

This evolution begins from infinite value of density $\bar{\rho}$
at some initial moment $\tau_0$. We can see here two different
variants for this beginning. For solutions with $w=-1{.}6$ and
$w=-1{.}8$ (we denote them as ``regular'' solutions) the scale
factor $a$ expands from $a=0$ like $a\sim\sqrt{\tau-\tau_0}$ at
the initial stage whereas the scale factor $b$ diminishes from
initial infinite value up to values $b\simeq b_0$ during some
percent of total lifetime of this universe. This behavior of
$b(\tau)$ looks like some variant of dynamical compactification,
because the parameter $b_0$ is arbitrary one in this model, we may
put $b_0$ to be sufficiently small.

Another type of evolution (``singular'' solutions) is represented
with dashed lines in Fig.~\ref{If1} for $w=-2$. For singular
solutions infinite value of density $\bar{\rho}$ at $\tau=\tau_0$
corresponds to nonzero value of the scale factor $a$ and $b=0$.
Obviously, these solutions are nonphysical and should be excluded.

All regular and singular solutions  in Fig.~\ref{If1} describe
accelerated expansion (for the factor $a$) at late stage of
evolution. Beginning of this stage may be seen in the graph of the
acceleration parameter $-q(\tau)$. Acceleration rate depends on
the parameters $w$, $\gamma$, $\Omega_0$ and the curvature
fraction (\ref{Omk}) $\Omega_k=-k_1(a_0H_0)^{-2}$ depending on the
sign $k_1$. If $\Omega_k\ne0$ ($k_1=\pm1$), one should use the
system (\ref{A3})~--~({\ref{B}}) instead of Eqs.~(\ref{sysA}),
(\ref{sysrho}). In this case we integrate numerically the function
$A'(\tau)$ simultaneously with solving the Cauchy problem for the
system (\ref{A3})~--~({\ref{B}}). We add here the natural initial
condition $A\big|_{\tau=1}=0$ to conditions (\ref{initial}).

For all reasonable values of four free parameters $w$, $\gamma$,
$\Omega_0$, $\Omega_k$ the stage of accelerated expansion appears
to be finite, because density $\bar{\rho}$ inevitably vanishes in
this model. In Fig.~\ref{If1} this effect may be seen in the
graphs $\bar{\rho}(\tau)$ with logarithmic scale in Y-direction.
We denote the moment of zero density by $\tau_*$:
$\bar{\rho}(\tau_*)=0$. For $\tau>\tau_*$ density $\bar{\rho}$
becomes negative and nonphysical, all energy conditions (in
particular, the weak energy condition) are violated.

This finite-time future singularity may be classified as the Type
IV singularity in accordance with the scheme from
Refs.~\cite{BambaCapNO12,NojOdinFR}. For this singularity
$a(\tau_*)$ is nonzero, $\bar{\rho}(\tau_*)$ equals zero, the
effective density and pressure
 $$
 \rho_{eff}=\frac{3H^2}{8\pi G}=\rho_c{A'}^2,\quad
p_{eff}=- \frac{2\dot H+3H^2}{8\pi G}=\frac{2q-1}3\rho_{eff}
 $$
 remain nonzero, but higher
derivatives of $H$ diverge at $\tau\to\tau_*$.

Note that the main features of the considered cosmological
solutions, in particular, the future singularity,  finite lifetime
$\tau_0\le\tau\le\tau_*$ and negative density for $\tau>\tau_*$
take place not only for $d=1$, but also for higher dimensions
$d\ge2$. In the case of $d\ge2$ additional dimensions after
substituting the components $G^\mu_\nu$ into Einstein equation
(\ref{Eeq}) and substitutions (\ref{tau}) in these  equations and
Eq.~(\ref{contin}) we have in the flat case $k_1=k_2=0$ the
following system \cite{PahwaChS}, generalizing  Eqs.~(\ref{B}) --
(\ref{sysrho}):
  \begin{eqnarray}
A''&=&\frac{
d(d-1)\,B'\big(\frac12B'-A'\big)-3(d+1)\,{A'}^2-3d\bar{p}_b}{d+2},\nonumber\\ 
\bar{\rho}'&=&-3\bar{\rho}A'-d(\bar{\rho}+\bar{p}_b)\,B',
\label{sysrhoD}\\
B'&=&\frac{\sqrt{3\big[(d+2)\,{A'}^2+2(d-1)\,
\bar{\rho}\big]/d}-3A'}{d-1}.\nonumber 
 \end{eqnarray}

\begin{figure}[th]
\includegraphics[scale=0.62,trim=5mm 1mm 2mm 0mm]{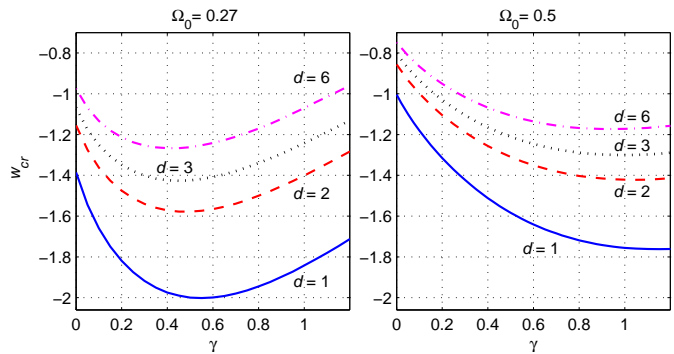}
\caption{Boundaries $w=w_{cr}$ between domains of regular (above)
and singular solutions (below a curve) for indicated values
$\Omega_0$ and $d$}
 \label{If2}\end{figure}

Solutions of the system (\ref{sysrhoD}) for $d\ge2$ were obtained
in Ref.~\cite{PahwaChS}, but some features of them were not
considered in that paper. For example, singular solutions with
nonzero value $a(\tau_0)$ (where $\bar{\rho}$ is infinite at the
initial moment $\tau_0$) also take place for $d\ge2$, if the value
$w$ is less than the critical value $w_{cr}(\gamma,\Omega_0)$. In
Fig.~\ref{If2} boundaries $w=w_{cr}$ separating domains of regular
and singular solutions on the $\gamma,\,w$ plane are presented for
different $d$ and $\Omega_0$. Singular solutions are described by
the inequality  $w<w_{cr}(\gamma,\Omega_0)$ and lie below
corresponding lines in Fig.~\ref{If2}.

Another important property of these cosmological solutions is
their finite-time future singularity, in other words,
inevitability of ``the end of the world'' because of vanishing
density at $\tau=\tau_*$ for all $d$ (see Fig.~\ref{If4} below).
The authors of Ref.~\cite{PahwaChS} did not pay attention to this
phenomenon, essential for their model. It is connected with the
chosen equation of state (\ref{pbro}) for pressure $\bar{p}_b$ in
extra dimensions. This drawback will be  eliminated with modifying
the model \cite{PahwaChS} in Sect.~\ref{Modif} after application
this model to describing observational data for Type Ia supernovae
in the next section.

\section{Application to supernovae observations}\label{Obs}

To apply the model to describing the observational data it is
convenient, following the authors of \cite{PahwaChS}, to use
Internet table \cite{SuperTable} for Type Ia supernovae in distant
galaxies. At the present moment  this updated table contains
redshifts $z=z_i$, distance moduli $\mu_i$ and errors $\sigma_i$
of $\mu_i$ for $N=580$ supernovae.

Redshift
\begin{equation}
 z=\frac{a_0}{a(t)}-1=e^{-A(\tau)}-1
  \label{z} \end{equation}
is associated with the value of $a$ at the time $t$ of a supernova
light emission. The distance modulus $\mu$ is the logarithmic
function
 $$
\mu=5\log\frac{D_L}{10\,\,\mbox{pc}},
 $$
 of the luminosity distance \cite{Clifton,PahwaChS}:
\begin{equation}
 D_L=(1+z)\int\limits_z^0\!\frac{d\tilde z}{H(\tilde z)}
 =\frac{a_0^2}{H_0a(\tau)}\,\int\limits_\tau^1\!\frac{d\tilde\tau}{a(\tilde\tau)}.
 \label{muDL} \end{equation}

To describe the data \cite{SuperTable} of Type Ia supernovae, for
given values $d$, $w$, $\gamma$, $\Omega_0$ of this model we
consider evolution of the scale factor $a(\tau)$ and dependence of
the numerical integral (\ref{muDL}) $D_L$ and $\mu$ on $\tau$. For
each value of redshift $z_i$ in the table \cite{SuperTable} we
calculate the corresponding $\tau=\tau_i$ with using Eq.~(\ref{z})
and linear approximation and the theoretical value
$\mu_{th}=\mu(\tau_i)$ for $\tau_i$ from Eq.~(\ref{muDL}). The
measure of differences between these theoretical values
$\mu_{th}=\mu_{th}(d,w,\gamma,\Omega_0,\Omega_k,z_i)$ and the
measured  values $\mu_i$ is \cite{PahwaChS}:
\begin{equation}
 \chi^2(d,w,\gamma,\Omega_0,\Omega_k)=\sum_{i=1}^N
 \frac{\big[\mu_i-\mu_{th}(d,\dots,z_i)\big]^2}{\sigma_i^2}.
  \label{chi} \end{equation}

The authors of Ref.~\cite{PahwaChS} calculated optimal parameters
$w$ and $\gamma$, minimizing the function (\ref{chi}) for the flat
model ($k_1=0$) with fixed $\Omega_0=0{.}27$ (\ref{Omega0}) and
$d\ge2$. In this approach for each $d\ge2$ they minimized the
function $\chi^2(w,\gamma)$ of two variables.

\begin{figure}[th]
\includegraphics[scale=0.62,trim=5mm 2mm 2mm 2mm]{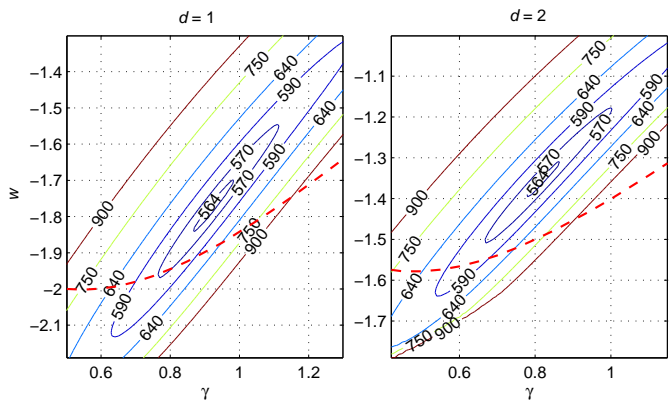}
\caption{Level lines of $\chi^2(w,\gamma)$ for $k_1=0$,
$\Omega_0=0{.}27$. The dashed line is the boundary of singular
solutions}
 \label{If3}\end{figure}

We generalize their approach to the case $d=1$ additional
dimension.  At the first step we fix $k_1=0$, $\Omega_0=0{.}27$ in
according with Ref.~\cite{PahwaChS} and obtain the picture of
level lines for the function $\chi^2(w,\gamma)$, presented in
Fig.~\ref{If3} for $d=1$ and $d=2$.

Here the dashed line is taken from Fig.~\ref{If2} and separates
regular and singular solutions. We see that for $d=1$ and $d=2$
the minimum of $\chi^2$ lies above this line, that is in the
domain of regular solutions. The same picture also takes place for
$d\ge3$.

For each $d\ge1$ we calculated minimums for the function of two
variables $\chi^2(w,\gamma)$ and coordinates $w$, $\gamma$ of this
minimum. They are represented in Table~\ref{Tab1}.

\begin{table}[th]
\caption{Minimum of $\chi^2$ and optimal values $w$, $\gamma$ for
fixed $k_1=0$, $\Omega_0=0{.}27$ and various $d$}
\begin{tabular}{||c||c|c|c|c|c||c||}  \hline
 $d$       &     1  &    2   &    3   &       6  &    10 & {\footnotesize $\Lambda$CDM}\\ \hline
 {\footnotesize min\,}$\chi^2$&563.15 & 563.41 & 563.52 &  563.65& 563.7&  563.29\\ \hline
 $w$       & $-1.77$&$-1.355$&$-1.212$&$-1.067$&$-1.01$& -\\ \hline
$\gamma$   &  0.925 &0{.}819 & 0{.}772&0{.}715&0{.}686& -\\
\hline
 \end{tabular}
 \label{Tab1}\end{table}

We compare these  minimal values with the value of the function
(\ref{chi}) for the flat $\Lambda$CDM model with the same
parameters $k_1=0$, $\Omega_m=\Omega_0=0{.}27$ (therefore,
$\Omega_\Lambda=0{.}73$) and  the same supernova data
\cite{SuperTable}. We see that the predictions are rather close,
and for $d=1$ the model \cite{PahwaChS} fits the data better than
the flat $\Lambda$CDM model.

At the next step for more precise estimation of optimal model
parameters we consider variations of fractions $\Omega_0=\Omega_m$
and $\Omega_k$ for matter density and curvature respectively. One
should take into account these degrees of freedom in both models:
the model \cite{PahwaChS} and $\Lambda$CDM. In the model
\cite{PahwaChS} for each $d\ge1$ we minimize the function
(\ref{chi}) of four variables:
$\chi^2(w,\gamma,\Omega_0,\Omega_k)$. We also compare this results
with the same value of the $\Lambda$CDM model (where $\chi^2$
depends on $\Omega_m$ and $\Omega_k$) and keep in mind the
constraints on these parameters due to cosmic microwave background
anisotropy, galaxy clustering and other factors \cite{WMAP}:
 \begin{equation}
 \Omega_m=0.2743\pm0.0072,\quad -0.0133< \Omega_k< 0.0084.
  \label{limOm} \end{equation}

 Numerical search of this
minimum includes a starting point (for example, the values from
Table~\ref{Tab1}), analysis of gradients or increments for
$\chi^2$ and  the constraints (\ref{limOm}). The results of
calculation with optimal values of the model parameters are
presented in Table~\ref{Tab2}.

\begin{table}[bh]
\caption{Minimum of $\chi^2$ and optimal values $w$, $\gamma$,
$\Omega_0$, $\Omega_k$}
\begin{tabular}{||c||c|c|c|c|c||c||}  \hline
 $d$       &     1  &    2   &    3   &      6  &    10 & {\small $\Lambda$CDM}\\ \hline
 {\small min\,}$\chi^2$&563.136 & 563.39 & 563.506 &563.634& 563.69&  563.058\\ \hline
 $w$       & $-1{.}740$&$-1.323$&$-1.2032$&$-1.061$&$-1.003$& -\\ \hline
$\gamma$   &  0{.}926 &0{.}821 & 0{.}7739&0{.}7174&0{.}696& -\\
\hline
$\Omega_0$   &  0{.}2815 &0{.}279 & 0{.}274  &0{.}2673 &0{.}267&0.2716\\
\hline
$\Omega_k$   &  0{.}0084 &0{.}0084&0{.}0084&0{.}0084&0{.}0084&$-0.0133$\\
\hline
 \end{tabular}
 \label{Tab2}\end{table}

We see that the $\Lambda$CDM model is more sensitive to variations
of  $\Omega_m$ and $\Omega_k$ and the better result for this model
is achieved. Here optimal values of the model parameters are
determined by the constraints (\ref{limOm}). We impose these
constraints on the model \cite{PahwaChS} though they are not
strictly applicable to it. In this model  min\,$\chi^2$ weakly
depends on $\Omega_0$ and $\Omega_k$, so we can not diminish
$\chi^2$ appreciably if we slightly broaden the limitations
(\ref{limOm}).

\begin{figure}[th]
\includegraphics[scale=0.8,trim=2mm 5mm 2mm 0mm]{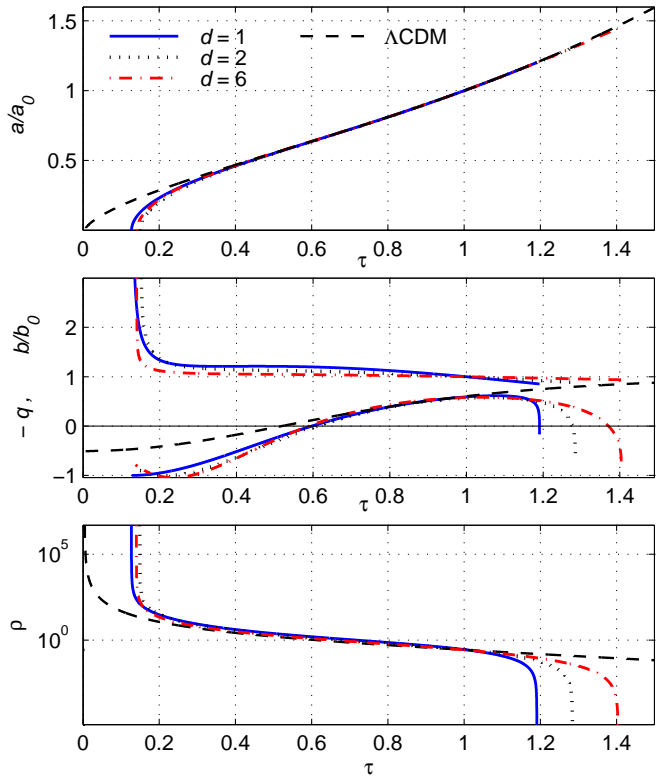}
\caption{Scale factors $a(\tau)$, $b(\tau)$, acceleration
parameter $-q(\tau)$ and density $\bar\rho(\tau)$ for the optimal
solutions from Table~\ref{Tab2}}
 \label{If4}\end{figure}

In Fig.~\ref{If4} one can see evolution of the scale factor
$a(\tau)$, (and $b$ for the model \cite{PahwaChS}), the
acceleration parameter $-q(\tau)$ and density $\bar\rho(\tau)$ for
the $\Lambda$CDM model and the model \cite{PahwaChS} with
 $d=1$ (solid lines), $d=2$ (dots) and $d=6$
(dash-dotted lines). For all these models we use the optimal
parameters from Table~\ref{Tab2}.

Evolution of the scale factor $a(\tau)$ for the model
\cite{PahwaChS} with different $d$ and for the $\Lambda$CDM model
is very close up to $z\simeq1{.}5$ ($a>0{.}4a_0$), before this
epoch the $\Lambda$CDM model demonstrates slower expansion. This
difference is more visible for the acceleration graphs $-q(\tau)$.
The scale factor $b$ for the case \cite{PahwaChS} diminishes to
$b\simeq b_0$ according to the mentioned above compactification
scheme (compare with the regular solutions in Fig.~\ref{If1}).

Behavior of cosmological solutions in the future for both models
is also different. The $\Lambda$CDM model demonstrates unlimited
accelerated expansion whereas for the model \cite{PahwaChS} the
acceleration turns into deceleration and inevitability results in
the above mentioned zero density $\bar{\rho}$ at $\tau=\tau_*$
with nonphysical values $\bar{\rho}<0$ for $\tau>\tau_*$. The
finite lifetime of this universe depends on $d$, it is the
smallest for $d=1$. In the next section we discuss how to
eliminate this essential drawback of the model.

\section{Modification of the model}\label{Modif}

We have noted that all cosmological solutions in the model
\cite{PahwaChS} have the finite-time future singularity. This
inevitable ``end of the world''  is connected with the chosen
power law dependence (\ref{pbro}) of pressure $\bar{p}_b$ in extra
dimensions on density $\bar\rho$. The terms with the factor
$\bar{p}_b$ in equations (\ref{sysrho}) for $d=1$ or
(\ref{sysrhoD}) for $d>1$ determine rate of density decreasing
when $\bar\rho$ is small at the end of its evolution. In this case
the leading terms in the mentioned equations are
 \begin{equation}
 \bar{\rho}'\simeq\left\{\begin{array}{ll}\bar{p}_bA',& d=1,\\
 -d\bar{p}_bB',& d>1,\end{array}\right.\qquad
\bar{\rho}\to0.
 \label{rhopr} \end{equation}
 For $\bar{\rho}\to0$ we have nonzero values $A'$ and $B'$, so for
the weak power law dependence (\ref{pbro})  the approximate
equation (\ref{rhopr}) $\bar{\rho}'\simeq-C{\bar\rho}^{1-\gamma}$
has the finite solution
 \begin{equation}
 \bar\rho\simeq\big[\gamma
C(\tau_*-\tau)\big]^{1/\gamma}.
 \label{rhof} \end{equation}

To avoid this finiteness we are to modify the equation of state
(power law dependence) (\ref{pbro}) of the model \cite{PahwaChS}
for small $\bar\rho$. In particular, a linear dependence for
$\bar\rho$ close to zero
 \begin{equation}
 \bar{p}_b=w_0\bar{\rho},\qquad \bar{\rho}\to0
 \label{pb0} \end{equation}
 ensures infinite evolution with positive density.

The linear law (\ref{pb0}) for all $\bar\rho$ does not describe
the observed accelerated expansion. For good agreement with
observations we are to search an equation of state
$\bar{p}_b(\bar\rho)$ with slower growth of $|\bar{p}_b|$ at high
$\bar\rho$ similar to Eq.~(\ref{pbro}). We suggest the appropriate
variant of this dependence
 \begin{equation}
 \bar{p}_b=\bigg(w_1+\frac w{\rho_0+\bar{\rho}}\bigg)\bar{\rho}
 \label{pbm} \end{equation}
 with the linear law (\ref{pb0}) for $\bar\rho\ll\rho_0$ (here $w_0=w_1+w/\rho_0$)
and another linear law $\bar{p}_b\simeq w_1\bar{\rho}$ for
 $\bar\rho\gg\rho_0$.

\begin{figure}[th]
\includegraphics[scale=0.8,trim=2mm 5mm 0mm 0mm]{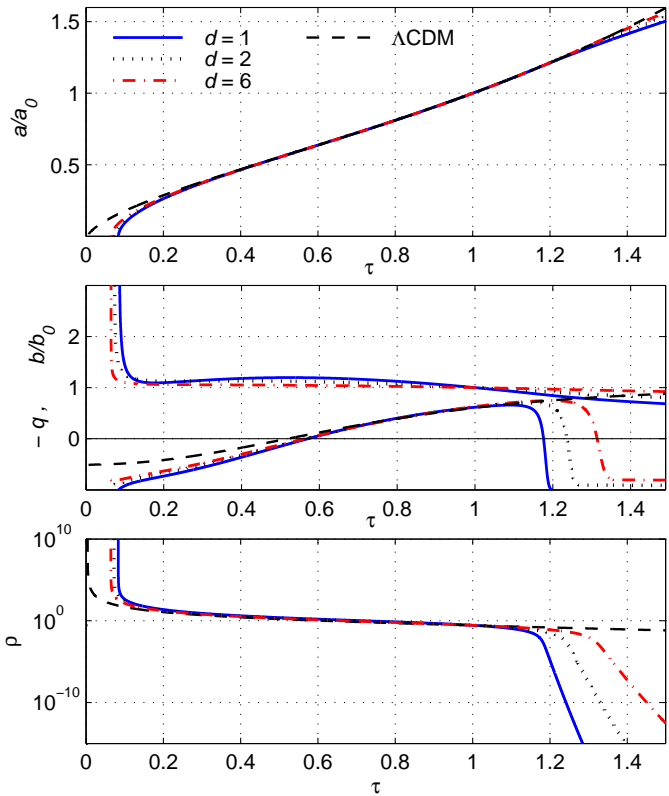}
\caption{Scale factors $a(\tau)$, $b(\tau)$, functions $-q(\tau)$
and $\bar\rho(\tau)$ for the model with Eq.~(\ref{pbm}) with the
parameters from Table~\ref{Tab3} }
 \label{If5}\end{figure}

The model (\ref{A3}) -- (\ref{B}) or (\ref{sysrhoD}) for $d>1$
with the linear-fractional equation of state (\ref{pbm}) makes it
possible to avoid finite lifetime of the type (\ref{rhof}) and to
transform it into the exponential asymptotic behavior
 \begin{equation}
 \bar\rho\sim\exp(-C\tau),\qquad C=\mbox{const}\cdot\Big(w_1+\frac
 w{\rho_0}\Big).
 \label{rhoex} \end{equation}
 This behavior results from the equation
$\bar{\rho}'\simeq-C{\bar\rho}$ and may be observed in graphs
$\bar{\rho}(\tau)$ in Fig.~\ref{If5}.

For the model with Eq.~(\ref{pbm}) we can find optimal values of
parameters $w$, $w_1$, $\rho_0$, $\Omega_0$, $\Omega_k$ presented
in Table~\ref{Tab3} and achieve better agreement with the
supernovae data \cite{SuperTable} than for the models $\Lambda$CDM
and \cite{PahwaChS} with Eq.~(\ref{pbro}). Cosmological solutions
for the model with Eq.~(\ref{pbm}) and parameters from
Table~\ref{Tab3} are shown in Fig.~\ref{If5}.
\begin{table}[hb]
\caption{Optimal parameters for the model with Eq.~(\ref{pbm}),
$\rho_0=0{.}005$ is
fixed 
 }
\begin{tabular}{||c||c|c|c|c|c||c||}  \hline
 $d$       &     1  &    2   &    3   &      6  &    10 & {\small $\Lambda$CDM}\\ \hline
 {\small min\,}$\chi^2$&562.898 & 562.814 & 562.79 &562.766& 562.757&  563.058\\ \hline
 $w$       & $-1.603$&$-1.026$&$-0.84$&$-0.658$&$-0.587$& -\\ \hline
$w_1$   & $-0{.}195$ &$-0.343$& $-0{.}387$&$-0{.}426$&$-0{.}439$& -\\
\hline
$\Omega_0$   & \multicolumn{5}{c||}{$0{.}2815$} &0.2716\\
\hline
$\Omega_k$   &  \multicolumn{5}{c||}{$-0.0133$}&$-0.0133$\\
\hline
 \end{tabular}
 \label{Tab3}\end{table}

We see in Table~\ref{Tab3} that the accuracy of the model with
Eq.~(\ref{pbm}) increases ($\chi^2$ diminishes) for large $d$,
unlike in the case with Eq.~(\ref{pbro}) in Table~\ref{Tab2}. We
should note that the values $\chi^2$ in Table~\ref{Tab3} are not
absolutely minimal, because we fixed the parameter
$\rho_0=0{.}005$. It is interesting, that for all $d$ we can
achieve smaller values min\,$\chi^2$, if we take smaller values of
$\rho_0$. But if $\rho_0\to0$, the factor $C$ in the exponent
(\ref{rhoex}) tends to infinity, the density $\bar{\rho}$
decreases too rapidly and the picture of vanishing $\bar{\rho}$
looks like in the finite  case in  Fig.~\ref{If4}. So we put the
restriction $\rho_0\ge0{.}005$ to exclude this almost
instantaneous transition to the state with $\bar{\rho}\simeq0$.
Under this constraint we have the optimal value $\rho_0=0{.}005$
and also $\Omega_0=0{.}2815$, $\Omega_k=-0.0133$ for all $d$.

Fig.~\ref{If5} demonstrates cosmological solutions for the model
with Eq.~(\ref{pbm}) with the optimal values of parameters from
Table~\ref{Tab3}. For both models Eqs.~(\ref{pbm}) and
(\ref{pbro})  in Figs.~\ref{If4} and \ref{If5} the acceleration
epoch is finite and its duration depends on $d$ in the same
manner. But after this epoch for the model with Eq.~(\ref{pbm}) we
see here infinite decelerated expansion.

Graphs of $a(\tau)$, $-q(\tau)$ and $\bar\rho(\tau)$ here are more
close to the dashed lines for the $\Lambda$CDM model during the
acceleration epoch than the similar curves in Fig.~\ref{If4}. But
after the mentioned epoch predictions of the $\Lambda$CDM model
and the model (\ref{pbm}) sharply diverge.

\begin{figure}[th]
\includegraphics[scale=0.7,trim=5mm 1mm 2mm 0mm]{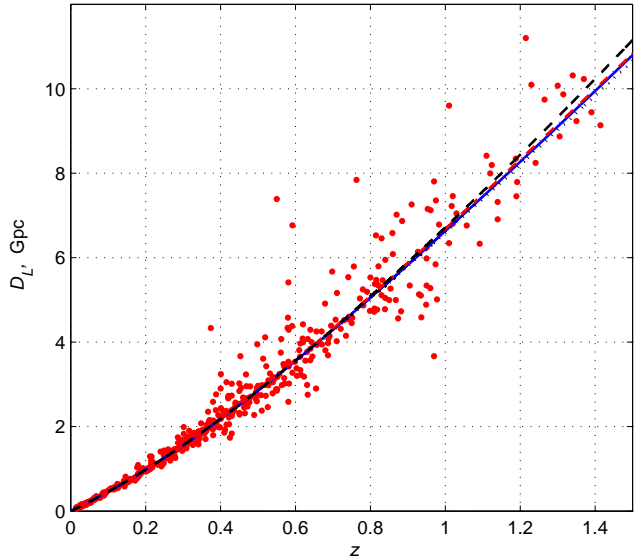}
\caption{Luminosity distance $D_L$ in Gpc depending on redshift
$z$ for the models $\Lambda$CDM and (\ref{pbm}) with parameters
from Table~\ref{Tab3}. Dots are the data \cite{SuperTable}}
 \label{If6}\end{figure}

In Fig.~\ref{If6} we present how the model with Eq.~(\ref{pbm})
for $d=1$, 2, 6 and the $\Lambda$CDM model describe the supernovae
data from the site \cite{SuperTable} (dots) in the $z$-$D_L$
plane. All these models were considered below in Fig.~\ref{If5},
we use the same optimal parameters from Table~\ref{Tab3} and the
same notations for the curves, in particular, the dashed line for
the $\Lambda$CDM model.

The presented curves are very close in the region $z<1$, for
larger $z$ the $\Lambda$CDM line slightly diverges from others.
These lines are result of optimal fitting to the observational
data \cite{SuperTable} (580 dots in Fig.~\ref{If6}). The values
$\chi^2$ in Table~\ref{Tab3} show rather good results for the
model \cite{PahwaChS} with Eq.~(\ref{pbm}) for pressure, but these
values are not the best fit, because we fixed $\rho_0$ to avoid
the mentioned above sharp transition to $\bar{\rho}\simeq0$.

\section{Conclusion}

The gravitational model of Pahwa, Choudhury and Seshadri
\cite{PahwaChS} with additional spatial dimensions and anisotropic
pressure provides accelerated expansion of the universe
corresponding to observational data for Type Ia supernovae
\cite{SuperTable} simultaneously with dynamical compactification
of $d$ extra dimensions. It is important that such a behavior of
solutions results from rather simple equations of state
(\ref{pbro}). This approach is more natural in comparison with the
scheme of Mohammedi \cite{Mohammedi02}, where complicated
equations of state are deduced from the constructed solutions, in
particular, from the solution (\ref{Moh}) described below.

The authors of Ref.~\cite{PahwaChS} did not consider the case
$d=1$, but we found that for the chosen in Ref.~\cite{PahwaChS}
power law equation of state (\ref{pbro}) it is the case $d=1$ that
yields the best fit (see Tables~\ref{Tab1}, \ref{Tab2}).

Unfortunately, the model \cite{PahwaChS} with the power law
dependence (\ref{pbro}) inevitably predicts the finite-time future
singularity of Type IV in classification  from
Refs.~\cite{BambaCapNO12,NojOdinFR}. It is connected with
vanishing density $\bar\rho$ at finite time $\tau=\tau_*$ (and
negative density for $\tau>\tau_*$). Evolution of this universe is
broken at $\tau_*$, the finite lifetime is shorter for small $d$
(see Fig.~\ref{If4}).
 We demonstrated in Sect.~\ref{Modif}, that this drawback has
technical nature. It is connected with too weak dependence
$\bar{p}_b(\bar\rho)$ for small $\bar\rho$ in the power law
equation of state (\ref{pbro}). If we modify this law and choose a
linear dependence (\ref{pb0}) for small $\bar\rho$, we obtain an
infinite cosmological evolution with positive density (but
vanishing at $\tau\to\infty$). We suggest the linear-fractional
variant (\ref{pbm}) of dependence $\bar{p}_b(\bar{\rho})$ to solve
the following two problems: (a) to avoid ``the end of the world''
of the type (\ref{rhof}) and  (b) to describe 580 Type Ia
supernovae data points from the site \cite{SuperTable}. The
dependence (\ref{pbm}) is a bit more complicated than
Eq.~(\ref{pbro}), but it successfully conserves positive density
$\bar\rho$ during infinite lifetime and fits the data
\cite{SuperTable} better than the $\Lambda$CDM model and the model
from Ref.~\cite{PahwaChS} with Eq.~(\ref{pbro}). Although the
simplicity of the equations of state (\ref{Pbrho1}) is the
important advantage of the model \cite{PahwaChS}, we are to step
back from this simplicity. But in our opinion, the dependence
(\ref{pbm}) is the minimal retreat that solves this problem.

Our calculations should be compared with predictions of other
multidimensional models
\cite{PahwaChS,Mohammedi02,Darabi03,Darabi09,BringmannEG03,PanigrahiZhCh06,MiddleSt11,FarajollahiA10,Neupane}.
We shall consider the models, describing the late time
acceleration of $a(t)$ together with a  contraction of $b(t)$, in
particular, the Mohammedi model in Ref.~\cite{Mohammedi02} with
the ansatz (\ref{Mohans}) $b/b_0=(a/a_0)^{-n}$.
 It ensures a dynamical compactification, if $n$ is positive and
 $a(t)$ expands. For $n$ satisfying the equality
  $$ dn(dn - n - 6)+6  = 0$$
  a set of solutions with accelerated expansion
 was obtained in  Ref.~\cite{Mohammedi02} in the form
 \begin{eqnarray}
 a/a_0 =  C_1\exp (\mu t)- C_2\exp (-\mu t)\nonumber \\
 =\tilde C_1\exp (\tilde\mu\tilde\tau)+(1-\tilde C_1)\exp
(-\tilde\mu\tilde\tau).
 \label{Moh} \end{eqnarray}
  Here $\tilde\tau=\tau-1$, the natural condition $a\big|_{t=t_0}=a_0$ must be saticfied.

We mentioned above, that the equation of state in the Mohammedi's
approach may be determined at the last stage after substitution of
the expressions (\ref{Moh}) and (\ref{Mohans}) into
Eqs.~(\ref{Eeq}) with a $\Lambda$ term.  In particular, the
relation between $P_a$ and $\rho$ for solutions (\ref{Moh})
results from the first two equations (\ref{Eeq}) (in our
notations)
 $$\begin{array}{c}
3k_1/a^2-\Lambda=8\pi G\rho,\\
\big[4C_1C_2\mu^2(2-2dn-dn^2)-k_1\big]/a^2\rule{0mm}{1.3em}\\
 -dn(n+1)\,\mu^2+\Lambda=8\pi
GP_a,\rule{0mm}{1.3em}
 \end{array}$$
if we exclude $a^2$. This equation of state is mush more
complicated than its analog $P_a=0$ for the model \cite{PahwaChS},
in addition it has the negative limit of $P_a$ at $a\to\infty$ for
the case $\Lambda=0$.

If we accept these complicated equations of state for the model
\cite{Mohammedi02}, we can obtain the optimal solution
(\ref{Moh}), minimizing the sum  $\chi^2$ (\ref{chi}) for the same
supernovae data \cite{SuperTable}. For this purpose we use 2
fitting parameters of these solutions: $\tilde C_1$ and
$\tilde\mu$. The calculations result in the optimal values $\tilde
C_1=1{.}229$, $\tilde\mu=0{.}679$ and the corresponding minimum
$\chi^2\simeq564{.}4$. This minimum is close to the results of the
considered model \cite{PahwaChS} in Tables~\ref{Tab2} and
\ref{Tab3}. So we may conclude that the solution (\ref{Moh})
describes the supernovae data \cite{SuperTable} rather
successfully. It is interesting that solutions close to
Eq.~(\ref{Moh}) appeared in  Refs.~\cite{Neupane} in the brane
model.

In Refs.~\cite{Darabi03,Darabi09} Darabi obtained exponential
solutions
$a=C_1\exp (\mu t)$ for the model with varying 
$\Lambda\sim a^{-m}$. Such a solution with one fitting parameter
is less adaptable in comparison with Eq.~(\ref{Moh}), the optimal
sum (\ref{chi}) in this case $\chi^2>955$.

Note that in this paper for the model  with Eq.~(\ref{pbm}) we
practically used only two fitting parameters $w$ and $w_1$. The
value $\rho_0$ was fixed because for very small $\rho_0$ we have
better fit, but the sharp downfall of $\bar\rho(\tau)$ to
$\bar\rho\simeq0$ looks like the mentioned ``end of the world''.
The parameters $\Omega_0$  and $\Omega_k$ influence on minimum of
$\chi^2$ rather weakly for the model  with Eq.~(\ref{pbm}). If we
fix, for example, $\Omega_0=0{.}27$  and $\Omega_k=0$, minimums
for $\chi^2$ will differ from results in Table~\ref{Tab3} less
then 0{.}01 for all $d$.

Cosmological solutions in the model with Eq.~(\ref{pbm}) are
divided into regular and singular ones similarly to solutions with
Eq.~(\ref{pbro}) shown in Fig.~\ref{If1}. However, Fig.~\ref{If5}
demonstrates that for the optimal values of parameters from
Table~\ref{Tab3} solutions with Eq.~(\ref{pbm}) are regular.

It is interesting that the model \cite{PahwaChS} with both
considered variants of dependence $\bar{p}_b$ on $\bar{\rho}$
(\ref{pbro}) and (\ref{pbm}) predicts finiteness of the
acceleration epoch.  Its duration depends on $d$ in the same
manner (compare Figs.~\ref{If4} and \ref{If5}) and then
acceleration sharply turns to deceleration. In the case
(\ref{pbro}) this evolution is broken at $\tau=\tau_*$ with
$\bar\rho(\tau_*)=0$, but for the model with Eq.~(\ref{pbm}) the
decelerated expansion is infinite and density $\bar\rho(\tau)$
tends to zero in the exponential form (\ref{rhoex}).


\begin{thebibliography}{22}


\bibitem{accPerl} S. Perlmutter {\it et al.}, 
{\it Astrophys. J.} {\bf517} (1999) 565,
 astro-ph/9812133.

\bibitem{Riess}  A. G. Riess {\it et al.}, 
{\it Astron. J.} {\bf 116} (1998) 1009, astro-ph/9805201.

\bibitem{WMAP} E. Komatsu {\it et al.}, 
{\it Astrophys. J. Suppl.}  {\bf 192} (2011) 18, arXiv:1001.4538
[astro-ph.CO].

\bibitem{SDSS} B. A. Reid {\it et al.}, 
{\it MNRAS} {\bf 404} (2010) 60, arXiv:0907.1659 [astro-ph.CO].

\bibitem{CopelandST06} E. J. Copeland, M. Sami and S. Tsujikawa, 
{\it Int. J. Mod. Phys. D} {\bf15} (2006) 1753, hep-th/0603057.

\bibitem{Clifton}  T. Clifton, P. G. Ferreira, A. Padilla and C. Skordis,
{\it Physics Reports} {\bf 513} (2012) 1, arXiv:1106.2476
[astro-ph.CO].

\bibitem{BambaCapNO12} K. Bamba, S. Capozziello, S. Nojiri and S.~D.~Odintsov,
{\it Astrophys. and Space Science} {\bf342} (2012) 155, 
 arXiv:1205.3421 [gr-qc].

\bibitem{Kunz12} M. Kunz, 
{\it Comptes rendus - Physique} {\bf13} (2012) 539,  
arXiv:1204.5482 [astro-ph.CO].

\bibitem{Sahdev} D. Sahdev, 
{\it Phys. Rev. D} {\bf 30} (1984) 2495.

\bibitem{PahwaChS} I. Pahwa, D. Choudhury and T. R. Seshadri, 
{\it J. of Cosmology and Astroparticle Phys.} {\bf 1109} (2011)
015, arxiv:1104.1925  [gr-qc].

\bibitem{Mohammedi02}  N. Mohammedi,
 {\it Phys. Rev. D} {\bf 65} (2002) 104018, hep-th/0202119.

\bibitem{Darabi03} F. Darabi,
 {\it Class. Quant. Grav.} {\bf 20} (2003) 3385,
gr-qc/0301075.

\bibitem{Darabi09} F. Darabi,
 arXiv:0902.1863 [gr-qc].

\bibitem{BringmannEG03}
T. Bringmann, M. Eriksson and M. Gustafsson,
 {\it Phys. Rev. D} {\bf 68} (2003)
063516, astro-ph/0303497.

\bibitem{PanigrahiZhCh06}
D. Panigrahi, Y. Z. Zhang and S. Chatterjee,
 {\it Int. J. Mod. Phys. A} {\bf21} (2006) 6491, gr-qc/0604079.

\bibitem{MiddleSt11}
C. A. Middleton  and E. Stanley,
 {\it Phys. Rev. D} {\bf 84} (2011) 085013,  arXiv:1107.1828 [gr-qc].

\bibitem{FarajollahiA10}
H. Farajollahi, H. Amiri,
 {\it Int. J. Mod. Phys. D} {\bf19} (2010) 1823, arXiv:1005.3140 [gr-qc].

\bibitem{Neupane}
I. P. Neupane,
 {\it Class. Quant. Grav.} {\bf 26} (2009) 195008,
 arXiv:0905.2774;
 {\it Int. J. Mod. Phys. D} {\bf19} (2010) 2281, arXiv:1004.0254 [gr-qc].

\bibitem{Qiang05}
Li-e Qiang, Y. Ma, M. Han and D. Yu,
 {\it Phys. Rev. D} {\bf71} (2005) 061501, gr-qc/0411066.

\bibitem{Leon10}
 J. Ponce de Leon,
 {\it JCAP} {\bf03} (2010) 030, arXiv:1001.1961 [gr-qc].

\bibitem{MaartensK}  R. Maartens and K. Koyama, 
{\it Living Rev. Rel.} {\bf 13} (2010) 5, arXiv:1004.3962
[hep-th].


\bibitem{KamenMP01} A. Y. Kamenshchik, U. Moschella and V. Pasquier,
{\it Phys. Lett. B} {\bf 511} (2001) 265, gr-qc/0103004.

\bibitem{GKamenMP03} V. Gorini, A. Y. Kamenshchik and U. Moschella, {\it Phys. Rev.
D} {\bf 67} (2003) 063509, astro-ph/0209395.

\bibitem{CaldwellDS98}
R. R. Caldwell, R. Dave and P. J. Steinhardt,
{\it Phys. Rev. Lett.} {\bf 80} (1998) 1582, astro-ph/9708069.

\bibitem{ArmenMukhS00} C. Armend\'ariz-Pic\'on, V. F. Mukhanov and P. J. Steinhardt,
{\it Phys. Rev. Lett.} {\bf 85} (2000) 4438, astro-ph/0004134.

\bibitem{Chameleon03} J. Khoury and A. Weltman, 
{\it Phys. Rev. D} {\bf 69} (2004) 044026, astro-ph/0309411.

\bibitem{SotiriouF}
T. P. Sotiriou and V. Faraoni, 
{\it Rev. of Modern Phys.} {\bf 82} (2010) 451, arXiv:0805.1726
[gr-qc].

\bibitem{NojOdinFR}  S. Nojiri and S. D. Odintsov,
{\it Phys. Rept.} {\bf 505} (2011) 59, 
arxiv:1011.0544 [gr-qc].



\bibitem{SuperTable} Supernova Cosmology Project, http://supernova.lbl.gov/Union/

\end{thebibliography}
\end{document}